\documentclass[11pt]{article}
\usepackage[final]{graphics}

\font\tenmsa=msam10
\font\sevenmsa=msam7
\font\fivemsa=msam5
\font\tenmsb=msbm10
\font\sevenmsb=msbm7
\font\fivemsb=msbm5
\newfam\msafam
\newfam\msbfam
\textfont\msafam=\tenmsa  \scriptfont\msafam=\sevenmsa
\scriptscriptfont\msafam=\fivemsa
\textfont\msbfam=\tenmsb  \scriptfont\msbfam=\sevenmsb
\scriptscriptfont\msbfam=\fivemsb

\global\mathchardef\lesssim "142E

\newcommand{\slL}{\raise.15ex\hbox{$/$}\kern-.53em\hbox{$L$}}
\newcommand{\slP}{\raise.15ex\hbox{$/$}\kern-.53em\hbox{$P$}}
\newcommand{\slR}{\raise.15ex\hbox{$/$}\kern-.53em\hbox{$R$}}
\newcommand{\slQ}{\raise.15ex\hbox{$/$}\kern-.53em\hbox{$Q$}}
\newcommand{\slK}{\raise.15ex\hbox{$/$}\kern-.53em\hbox{$K$}}
\newcommand{\slSigma}{\raise.15ex\hbox{$/$}\kern-.53em\hbox{$\Sigma$}}
\newcommand{\slcalP}{\raise.15ex\hbox{$/$}\kern-.63em\hbox{$\cal P$}}

\newcommand{\be}{\begin{equation}}
\newcommand{\ee}{\end{equation}}     
\newcommand{\bea}{\begin{eqnarray}}
\newcommand{\ena}{\end{eqnarray}}

\def\build#1\over#2{\mathrel{\mathop{\kern 0pt#1}\limits_{#2}}}

\font\tenimbf=cmmib10 at 12pt
\font\sevenimbf=cmmib10 at 7pt
\font\fiveimbf=cmmib10 at 5pt
\newfam\imbf
\textfont\imbf=\tenimbf
\scriptfont\imbf=\sevenimbf
\scriptscriptfont\imbf=\fiveimbf
\def\imb{\fam\imbf\tenimbf}

\begin{document}
\begin{titlepage}
\title{\begin{center}
{\Huge LAPTH}
\end{center}
\vspace{5 mm}
\hrule
\vspace{20mm}
\bf{Two loop Compton and annihilation processes in thermal QCD\\
 }}
\author{
P.~Aurenche$^{(1)}$, F.~Gelis$^{(1)}$, 
R.~Kobes$^{(2)}$, H.~Zaraket$^{(1)}$}
\maketitle

\begin{center}
\begin{enumerate}
\item Laboratoire de Physique Th\'eorique LAPTH,\\
URA 1436 du CNRS, associ\'ee \`a l'Universit\'e de Savoie,\\
BP110, F-74941, Annecy le Vieux Cedex, France
\item Physics Department and Winnipeg Institute
for Theoretical Physics,\\
University of Winnipeg,
Winnipeg, Manitoba R3B 2E9, Canada
\end{enumerate}
\end{center}

\begin{abstract}
  We calculate the Compton and annihilation production of a soft
  static lepton pair in a quark-gluon plasma in the two-loop
  approximation. We work in the context of the effective perturbative
  expansion based on the resummation of hard thermal loops. Double
  counting is avoided by subtracting appropriate counterterms. It is
  found that the two-loop diagrams give contributions of the same
  order as the one-loop diagram. Furthermore, these contributions are
  necessary to obtain agreement with the naive perturbative expansion
  in the limit of vanishing thermal masses.
\end{abstract}
   \vskip 4mm
\centerline{\hfill LAPTH--719/99,  WIN--99/02,  hep-ph/9903307}
\vfill
\thispagestyle{empty}
\end{titlepage}
\section{Introduction} 

We consider the production of a soft static lepton pair in a quark
gluon plasma. In our approach we follow strictly the hard thermal loop
(HTL) scheme of \cite{BraatP1,FrenkT1}, appropriate for a plasma in
equilibrium at high temperature with a small coupling constant between
the quarks and the gluons in the plasma. We thus construct the loop
expansion using effective vertices and propagators instead of bare
ones. The production rate of static virtual photons has already been
evaluated at the one-loop level in the effective theory
\cite{BraatPY1}. Previous works \cite{AurenGKP2,AurenGKZ1} have shown
that two-loop diagrams are of equal importance as the one-loop
diagram.  These large contributions are associated with new processes
(namely bremsstrahlung) which arise only at the two--loop level. However,
these processes are only a part of the full two-loop diagrams.  In
this paper we extend our previous work \cite{AurenGKZ1} and complete
the calculation of these diagrams. The underlying physical processes
to be considered are Compton and quark-antiquark annihilation which
are already present at one loop. The underlying
physical processes indicates that there is a possible double counting
between one and two-loop results. More precisely, the two-loop
diagrams of Fig. 1 are already contained in the one-loop diagram of
\cite{BraatPY1} (with effective propagators and vertices) when the
gluon is hard and time-like. We take care of this problem by
subtracting the appropriate counterterms.

We show that the two-loop contribution corrects in a crucial way the
calculation of the Compton and annihilation processes based on the
one-loop approximation \cite{Zarak2,Gelis7}. Adding one and two loop
contributions we find that in the limit of vanishing thermal masses we
recover the result already found in \cite{AltheA1} where the authors
used the bare theory.  This solves a long standing puzzle related to
the fact that the one-loop approximation in the effective theory did
not reduce to known results in the naive perturbative expansion
\cite{Braat3}.

It is not possible to derive analytically the full leading expression,
in powers of $g$, of the rate of static virtual photon production up
to two loops in the effective theory. However, it is relatively easy to
obtain analytically the large logarithmic $\ln(1/g)$ terms.
After some general considerations concerning our approach and the
approximations useful to extract the leading logarithmic behavior we
review the one-loop results and then discuss in Sec.~\ref{sec:2loop}
the two-loop calculation. The next section is devoted to an
approximate evaluation of the counterterms. Combining everything we then
obtain in Sec.~\ref{sec:tot2loop} the rate of virtual photon
production. In the Appendix we give the exact result for the
counterterms, using effective vertices, where we show that the leading
logarithmic behavior is the same as that obtained using the
simplified version.

\section{General considerations}
\label{sect:gencons}
\begin{figure}[htbp]
  \centerline{
    \resizebox*{!}{2.1cm}{\includegraphics{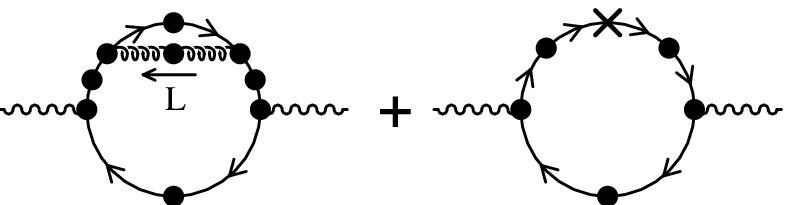}}
    }
  \vskip 2mm
  \centerline{
    \resizebox*{!}{2.1cm}{\includegraphics{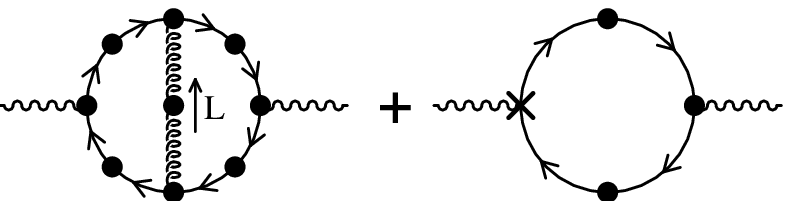}}
    }
  \caption{\footnotesize{Two-loop contributions. A black dot denotes an effective
      propagator or vertex. Crosses are HTL counterterms.}}
  \label{fig:2loop}
\end{figure}

It is well known that the production rate of a photon of invariant
mass $\sqrt{Q^2}$ decaying into a lepton pair is proportional to the
imaginary part of the retarded vacuum polarization of the considered
photon \cite{Weldo3,GaleK1}:
\begin{equation}
  {{dN}\over{dtd^3{\imb x}}}=-
  {{dq_0d^3{\imb q}}\over{12\pi^4}}\;
  {\alpha\over{Q^2}}\,n_{_{B}}(q_0)\,
  {\rm Im}\,\Pi^{^{RA}}{}_\mu{}^\mu(q_0,{\imb q})\; .
  \label{virtualphot}
\end{equation}
At the one-loop order, the trace of the vacuum polarization tensor is
given by the fermion loop with effective propagators and
vertices, as calculated in \cite{BraatPY1}. The two-loop
diagrams with their associated counterterms are shown in
Fig.~\ref{fig:2loop}. The justification of the use of counterterms is obvious.
For example, when evaluating the first diagram in Fig. 1, one has to
integrate over a region of phase-space where the gluon momentum $L$
can be time-like ($L^2 \ge 0$) and hard ($L^\mu \sim T$). This
contribution is already included when using effective fermion
propagators in the one-loop diagram. The role of the self-energy
counterterm is to remove from the two-loop contribution the part which
is already included at one loop. A similar discussion can be made
concerning the diagrams on the second line of Fig.~\ref{fig:2loop},
which indicate that the vertex counterterm
must be introduced.

More formally, the counterterms automatically arise when we
construct the perturbative expansion with effective propagators and
vertices rather than with the original propagators and vertices. The
effective perturbative expansion is nothing but a reorganization of
the usual perturbative expansion. The effective Lagrangian ${\cal
  L}_{\rm eff}$, which leads to effective propagators and vertices,
already includes some one-loop thermal contributions. In order to
keep the same theory as that given by the original Lagrangian ${\cal
  L}$ one has to supplement the effective Lagrangian with counterterms
to subtract order by order the loop-corrections already included in
the effective Lagrangian so that
\begin{equation}
{\cal L} = {\cal L}_{\rm eff} + {\cal L}_{\rm c.t.}.
\end{equation}
The counterterms are then treated perturbatively and thus appear in
the expansion when looking at higher order topologies.

\begin{figure}[htbp]
  \centerline{
    \resizebox*{!}{2.1cm}{\includegraphics{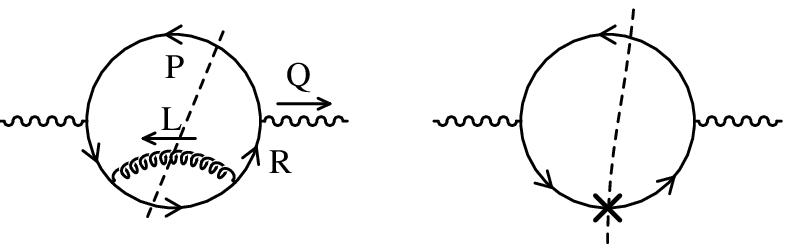}}
    }
  \vskip 2mm
  \centerline{
    \resizebox*{!}{2.1cm}{\includegraphics{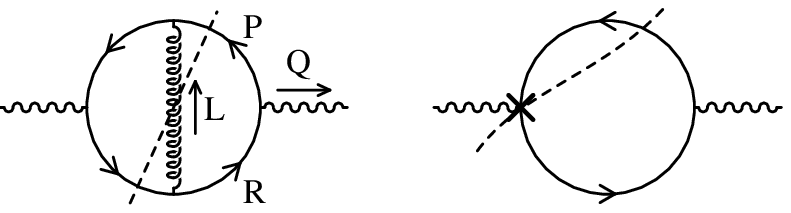}}
    }
  \caption{\footnotesize{Two-loop contributions in the hard momenta
      limit. Lines without any dressing denote the hard limit of
      effective propagators, in which we keep a thermal mass.}}
  \label{fig:2loopsm}
\end{figure}

The imaginary part of the retarded vacuum polarization in
Eq.~(\ref{virtualphot}) can be expressed as a sum over different cuts
through the diagrams of Fig.~\ref{fig:2loop}. Calculating the cut
diagrams with the full complexity of effective propagators and
vertices would be a formidable task. To extract only the leading
behavior of the two-loop contributions one can make some useful
simplifications. This leading contribution arises essentially when a
hard momentum flows in the quark loop. This is a consequence of the
phase space available, $\int d^4 p \sim T^4$ in the hard region,
compared to $g^4 T^4$ for the soft region. Besides that, for soft
fermion momentum $P$, the leading contribution should be contained
entirely in the one-loop diagram since effective propagators and
vertices give the complete behavior at scale $g T$. In the
soft quark region, the two-loop diagrams (with associated
counterterms) will only give sub-dominant terms while new dominant
contributions can occur in the hard quark regime. Concentrating now on
the region where the quarks are hard, it is sufficient to use bare
vertices rather than effective ones and to keep only the time-like
sector of cut fermion propagators, which largely dominates over the
space-like sector. To see that, one can compare the behavior of the
spectral density $\delta (P^2) \slP ={\cal O}( 1/ p)$ in the time-like
region with its behavior $ g^2 T^2 \slP /p^4 $ in the space-like
region when the momentum $p$ is hard. We are therefore led to
calculate the graphs of Fig.~\ref{fig:2loopsm} with time-like cut
fermion lines.

With these simplifications, two types of cuts can be
distinguished: those going through the gluon propagator and those
going only through two fermion propagators. For the moment we ignore
the latter because kinematical constraints would require either $P$
and $R$ (see Fig. 2 for the notations) to be time-like and soft (which
would lead to a non leading contribution because of the small size of
the phase-space), or $P$ to be space-like and hard so that the
loop-corrections make these diagrams sub-leading compared to the
one-loop result. Cutting through the gluon line, it is convenient to
distinguish the space-like $L^2$ region from the time-like one. The
case of $L^2 < 0$ has been discussed in detail in \cite{AurenGKZ1}
where it has been shown to contain a new physical process
(bremsstrahlung) absent from the one-loop result.  No counterterms
were needed in this region.  From now on, we concentrate on the case
$L^2 > 0$.  Physically, this region corresponds to Compton and annihilation
processes, which are already included in an approximate way in the
one-loop calculation of Braaten, Pisarski and Yuan
(BPY)~\cite{BraatPY1}. We expect a better approximation of these
processes to come out from the evaluation of the two-loop diagrams for
the following reason: when evaluating these two-loop graphs, we do not
neglect $P$ and $R$ compared to $L$ in the self-energy and vertex
corrections since $P$ and $R$ are hard and comparable to $L$. On the
contrary, in the HTL approximation ``external" momenta $P$ and $R$ are
neglected compared to the loop-momentum $L$ in the resummed
self-energy corrections to the fermion propagators and in the estimate
of the loop correction leading to the effective vertex. We therefore
anticipate that in our two-loop calculations, terms of leading order
in $L$ in the matrix element will be compensated by counterterms while
terms of lower degree in $L$ will lead to new contributions.

The counterterm method described here is not the only way which can be
used to calculate the two-loop diagrams while avoiding double counting
of thermal corrections. Since, as mentioned above, effective
propagators and vertices do not have the correct behavior in the hard
limit (more precisely when some of their external momenta are hard and
space-like), an alternative way to proceed consists of introducing a
cutoff scale intermediate between $g T$ and $T$. In this method, one
uses effective propagators and vertices in loops carrying a momentum
below the cutoff. Above the cutoff, bare propagators and vertices can
be used, and a loop correction must be inserted on the hard
propagator.  When adding the soft and the hard contributions the
cutoff dependence should drop out \cite{Gelis7,Gelis9}. This method
has been pioneered by Braaten and Yuan in their calculation
\cite{BraatY1} of the rate of energy loss in a hot plasma, and used
later in the calculation of the rate of hard real photon production by
a quark-gluon plasma~\cite{BaierNNR1,KapusLS1}. However it was not
followed in \cite{BraatPY1} where the effective propagators were used
up to the hard momentum scale.

A remark is worth making concerning the use of the cutoff method in
\cite{BaierNNR1,KapusLS1}. Indeed, in these two papers, a bare gluon
propagator is used even if the cutoff does not constrain the gluon to
be hard. As a consequence of the misuse of the cutoff in this way,
\cite{BaierNNR1,KapusLS1} missed the important contribution coming
from bremsstrahlung \cite{AurenGKZ1}.


In the following we will obtain analytically the leading $\ln (1/g)$
behavior of the  vacuum polarization diagram. The argument of the
logarithm is the ratio of a hard scale over a soft scale of order $g T$.
Technically, this arises from integrals of type $\int {dp/ p}$ or
$\int {dl/ l}$. To extract this leading behavior one does not need
the full details of the effective propagators in the soft region. It is
sufficient to approximate the fermion and gluon dispersion relations by
their  asymptotic forms in the hard region: the error involves ratios of
soft scales which are of order $1$ compared to $\ln (1/g)$.
Furthermore, only the upper branches (quasi-particles) of the dispersion
relations will contribute because the lower branches (collective modes)
of both the fermion and the gluon decouple exponentially fast in the
hard region and therefore cannot contribute a leading logarithm
term. A last technical simplification consists in keeping a constant
mass, independent of the momentum, along the dispersion curves: thus we
have $m_{_F}^2 \sim m_{\rm g}^2 \sim g^2 T^2$, the exact expression of
the thermal masses being irrelevant for the logarithmic behavior.

\section{One-loop result}
The virtual photon production rate has been calculated at one loop in
the effective theory \cite{BraatPY1}. In their calculation, BPY
distinguished three types of terms depending on whether the pole
(time-like) part or the cut (space-like) part of the fermionic
spectral density is taken into account. We already extracted
in~\cite{AurenGKZ1} the leading logarithmic behavior of the cut-cut
contribution and we showed that it should be compared to the $L^2 <0$
part of the two-loop diagrams. We give now the remaining leading
logarithmic part associated to the pole-cut contribution (the
pole-pole term does not lead to a logarithm because hard time-like
momenta are kinematically suppressed) which describes the Compton and
the annihilation processes. From Eq.~(11) of \cite{BraatPY1}, taking
the hard momentum limit in the integrand, one can easily extract the
leading logarithm given by
\begin{eqnarray}
        \left. \frac{dN}{d^4xdq_0 d^3{\imb q}}\right|_{\rm 1loop}&\approx&
        \frac{\alpha^2}{12\pi^4}\Big(\sum_{f}e^2_f\Big)
        \left(\frac{NC_{_{F}}g^2T^2}{8}\right)
        \frac{1}{q_0^2}
      \ln\left(\frac{q_0T}{q_0^2+m_{_{F}}^2}\right)\ ,
\label{eq:1loop}
\end{eqnarray} 
where $N$ is the number of colors, $C_{_{F}}\equiv (N^2-1)/2N$,
$m_{_{F}}^2\equiv g^2C_{_{F}}T^2/8$ is the thermal mass of soft quarks
and $e_{f}$ is the electric charge of the quark of flavor $f$, in
units of electron charge.

For later use, it is convenient to translate this formula into an
expression for the photon polarization tensor:
\begin{equation}
 {\rm Im}\,\Pi^{^{RA}}{}_\mu{}^\mu(q_0,{\imb 0}) \approx
- \frac{NC_{_{F}}}{8} \frac{e^2g^2}{4\pi}
\Big(\sum_{f} e_{f}^2\Big) q_0 T \ln\left( \frac{q_0 T}{q_0^2+m_{_{F}}^2} \right)\;
.
\label{eq:polar1l}
\end{equation}

\section{The two-loop calculation}
\label{sec:2loop}
\subsection{General expression and the logarithmic behavior}
The two-loop expression (without counterterms) has been derived in
\cite{AurenGKZ1} under the same simplifying assumptions as those
used here (hard fermion momenta, no effective vertices). There is
was found
\begin{eqnarray}
 &&
  \!\!\!{\rm Im}\,\Pi^{^{RA}}
  {}_\mu{}^\mu(q_0,{\imb q}) = 
  - \frac{NC_{_{F}}}{2} e^{2}g^{2}
   \int{{d^4P}\over{(2\pi)^3}}\int{{d^4L}\over{(2\pi)^3}}
  [n_{_{F}}(r_{0})-n_{_{F}}(p_{0})]\nonumber\\
  &&\times\;
  [n_{_{B}}(l_{0})+n_{_{F}}(r_{0}+l_{0})]\delta(P^{2}-m_{_{F}}^{2})
  \delta((R+L)^2-m_{_{F}}^{2})\epsilon(p_{0})\epsilon(r_{0}+l_{0})\nonumber\\
  &&\times\;
  \sum\limits_{a=T,L}\rho_a(L) P^a_{\rho\sigma}(L)
    \left[{{{\rm Trace}{}^{\rho\sigma}{}_{|_{\rm vertex}}}
        \over{\overline{R}{}^2(\overline{P+L}){}^2}}
      +{{{\rm Trace}{}^{\rho\sigma}{}_{|_{\rm self}}}
        \over{\overline{R}{}^2
          \overline{R}{}^2}}\right]
\label{2loop}
\end{eqnarray}
where $e$ is the electric charge of the quark running in the loop
and where the notation
$\overline{R}\equiv(r_{0},\sqrt{r^{{2}}+m_{_{F}}^{{2}}}\hat{\imb
  r})$ includes the thermal mass shift on the fermion propagator:
\begin{equation}
{{\overline {\slR}} \over {\overline R}^2 \pm i r_0
 \varepsilon}
 = {{\overline {\slR}} \over R^2 - m_{_{F}}^2  \pm i r_0 \varepsilon}.
\end{equation}
After contracting over the transverse and longitudinal gluon projectors
 \cite{Weldo1,LandsW1},
\begin{eqnarray}
  &&P^{^{T}}_{\rho\sigma}(L)=
  g_{\rho\sigma}-U_\rho U_\sigma
  +{{(L_\rho-l_0U_\rho)(L_\sigma-l_0U_\sigma)}\over{l^2}}\\
  &&P^{^{L}}_{\rho\sigma}(L)=
  -P^{^{T}}_{\rho\sigma}(L)+g_{\rho\sigma}-{{L_\rho L_\sigma}\over{L^2}}
  \; ,
  \label{eq:projectors}
\end{eqnarray}
where $U\equiv(1,{\imb 0})$ is the 4-velocity of the plasma in its rest
frame we arrive at:
\begin{eqnarray}
  &&\!\!\sum\limits_{a=T,L}\rho_a(L) P^a_{\rho\sigma}(L)
  \left[{{{\rm Trace}{}^{\rho\sigma}{}_{|_{\rm vertex}}}
      \over{\overline{R}{}^2(\overline{P+L}){}^2}}
    +{{{\rm Trace}{}^{\rho\sigma}{}_{|_{\rm self}}}
      \over{\overline{R}{}^2\overline{R}{}^2}}\right]
  = \sum\limits_{a=T,L}\rho_a(L) |{\cal M}_a|^2\nonumber\\
  &&\approx-4\left[\left(\rho_{_{T}}(L)-\rho_{_{L}}(L)\right)
    {{4p^2(\cos^2\theta^\prime-1)}\over{\overline{R}{}^2(\overline{P+L}){}^2}}
    \left(L^2-2{{Q^2(Q\cdot L)^2}\over{\overline{R}{}^2(\overline{P+L}){}^2}}\right)\right.\nonumber\\
  &&\qquad+2{{(Q+L)^2}\over{\overline{R}{}^2(\overline{P+L}){}^2}}
  \left(Q^2\rho_{_{L}}(L)+L^2\rho_{_{T}}(L)\right)\nonumber\\
  &&\qquad-2\rho_{_{T}}(L)\left(1-2{{(Q\cdot
  L)^2}\over{\overline{R}^2(\overline{P+L}){}^2}}
      \left.+{{Q^2L^2}\over 2}
          \left[{1\over{(\overline{R}{}^2)^2}}+{1\over{((\overline{P+L}){}^2)^2}}
      \right]\right)
      \right]
 \label{eq:static1}
\end{eqnarray}
where $\theta'$ is the angle between $\imb r$ and $\imb l$ (or $\imb
p$ and $\imb l$ since the virtual photon is static).  Unlike in our
previous work which dealt with the contribution of the space-like part
of the gluon phase space, we consider here the time-like part and use:
\begin{equation}
\rho_a(L)=2\pi\epsilon(l_{0})\delta(L^2-{\rm Re}\,\Pi_{a}).
\label{realpi}
\end{equation}

Eq.~(\ref{2loop}) accounts for the cuts shown in
Fig.~\ref{fig:2loopsm}. We still have to include the symmetric cut for
the vertex diagram and also the diagram with the self-energy correction
on the lower line. This will be done simply by multiplying the final
result by an overall factor of 2.

Since the external photon is massive, $Q^2 \sim g^2 T^2$, there does
not appear any collinear singularities of the type discussed in
\cite{AurenGKP2} and the logarithmic behavior arises only from terms
like $\int {dp/p}$ or $\int {dl/l}$. A simple power counting will then
allow us to isolate the relevant terms in Eqs.~(\ref{2loop}) and
(\ref{eq:static1}). We can use the following rules
\begin{itemize}
\item{} $\int dp_0\,\delta(P^{2}-m_{_{F}}^2)\sim\displaystyle{\frac{1}{p}}\;$,
\item{} $\int dl_0\,\delta(L^{2}-{\rm Re}\,\Pi_{a})\displaystyle{\sim\frac{1}{l}}$,
\item{} $\int
d\cos\theta'\,\delta((R+L)^{2}-m_{_{F}}^{2})\sim\displaystyle{\frac{1}{pl}}$,
\item{} $n_{_{F}}(r_{0})-n_{_{F}}(p_{0})\sim\displaystyle{\frac{q_{0}}{T}}$
since $q_0 = r_0 -p_0$ is soft,
\item{} $n_{_{B}}(l_{0})+n_{_{F}}(r_{0}+l_{0})\sim\displaystyle{\frac{T}{l_{0}}}
\sim \displaystyle{\frac{T}{l}}$,
\end{itemize}
to write, ignoring any irrelevant angular integrations,
\begin{equation}
{\rm Im}\,\Pi^{^{RA}} {}_\mu{}^\mu \sim e^2 g^2 q_0 
\int d p {d l \over l} \sum\limits_{a=T,L} |{\cal M}_a |^2.
\end{equation}
Estimating ${\overline {R}}^2 \sim 2 p q_0$ and likewise for 
${(\overline {P+L})}^2$, we find that only the terms
\begin{equation}
 -2\rho_{_{T}}\left(
1-2{{(Q\cdot L)^2}
\over
{\overline{R}{}^2(\overline{P+L}){}^2}}\right)
=
-2\rho_{_T}
\left( 1 + Q\cdot L 
\left( {1 \over \overline{R}^2} + {1 \over (\overline{P+L}){}^2}
\right) \right)\; ,
\end{equation}
leading to
\begin{equation}
{\rm Im}\,\Pi^{^{RA}} {}_\mu{}^\mu \sim e^2 g^2 q_0 
\int d p {d l}\, \left( {\cal O}({1\over p})+{\cal O}({1\over l})\right)
\end{equation}
contribute to the leading logarithmic behavior. It has been checked
by explicit analysis that all the other terms do not have the proper
scaling behavior to give a logarithm.  It can be noted here that the
relevant terms for the present analysis are totally different from
those in the bremsstrahlung case ($L^2 < 0$).  This is related to the
different behavior of the gluon spectral density in the space-like
region and the time-like region for hard momentum.  More precisely, in
our calculation of the trace using the Feynman gauge, the leading
bremsstrahlung terms were all contained in the vertex corrections
whereas for Compton and annihilation they appear in the diagrams with
self-energy corrections.

To summarize, in order to obtain the leading logarithmic terms for positive 
$L^2$  it is sufficient to calculate
\begin{eqnarray}
&&
  \!\!\!{\rm Im}\,\Pi^{^{RA}}
  {}_\mu{}^\mu(q_0,{\imb 0}) \approx 
 4 NC_{_{F}}\ e^2g^2 
   \int{{d^4P}\over{(2\pi)^3}}\int{{d^4L}\over{(2\pi)^3}}
  [n_{_{F}}(r_{0})-n_{_{F}}(p_{0})]\nonumber\\
  &&\times\;
  [n_{_{B}}(l_{0})+n_{_{F}}(r_{0}+l_{0})]\delta(P^{2}-m_{_{F}}^{2})
  \delta((R+L)^2-m_{_{F}}^{2})\epsilon(p_{0})\epsilon(r_{0}+l_{0})\nonumber\\
  &&\times\;
  2\pi\epsilon(l_{0})\delta(L^2-{\rm Re}\,\Pi_{T})\left[ 1+\left(\frac{Q\cdot L}
  {\overline{R}{}^{2}} + 
  \frac{Q\cdot L}{(\overline{P+L}){}^{2}}\right)\right] \nonumber\\
  &&\qquad \qquad \qquad \ 
\equiv {\rm Im}\,\Pi_{1}(q_0,{\imb 0})+{\rm Im}\,\Pi_{2}(q_0,{\imb 0})
\label{eq:formula}
\end{eqnarray}
The above decomposition of ${\rm Im}\,\Pi$ into two parts is natural:
${\rm Im}\,\Pi_{1}$ (the term $1$ in the square brackets) is dominated
by hard $p$ while the logarithm comes from the $l$ integration; on the
other hand, ${\rm Im}\,\Pi_{2}$ (the terms proportional to $Q\cdot L$)
is dominated by hard $l$ and it has a logarithm in the $p$
integration, which indicates that we can make different approximations
in each of these parts. From the discussion in
section~\ref{sect:gencons}, we anticipate that ${\rm Im}\,\Pi_{1}$ is
a new two-loop contribution while ${\rm Im}\,\Pi_{2}$ should be
compensated by counterterms.

\subsection{Hard $p$ region (${\rm Im}\,\Pi_{1}$)}

\begin{figure}[t]
  \centerline{\resizebox*{!}{5cm}{\includegraphics{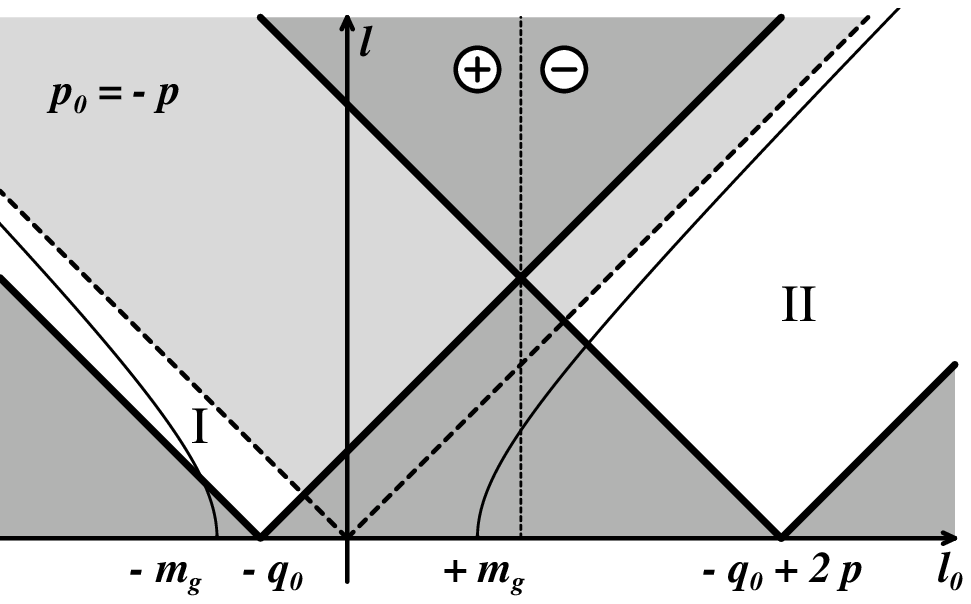}}}
  \centerline{\resizebox*{!}{5cm}{\includegraphics{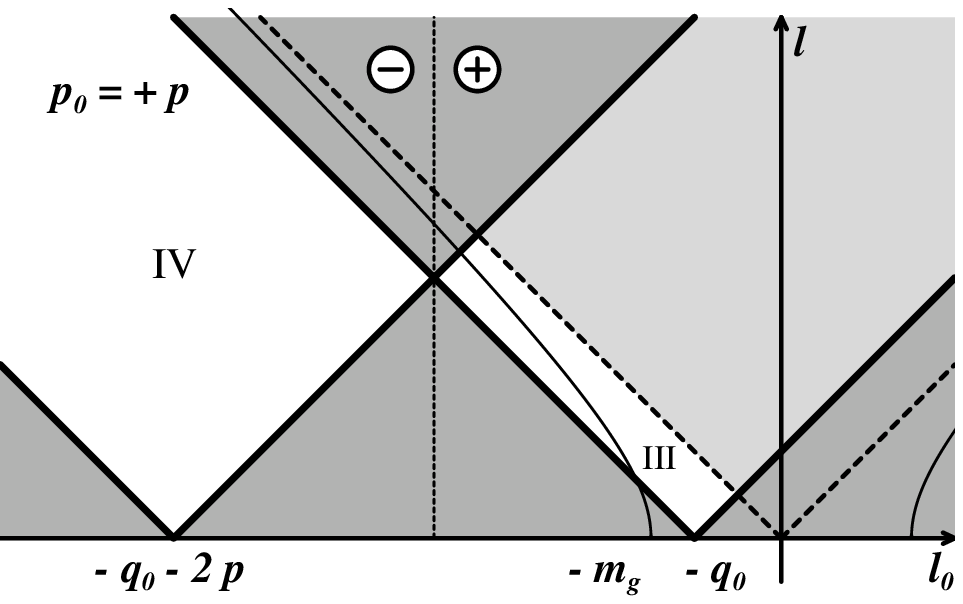}}}
  \caption{\footnotesize{Allowed domains in the $(l_0,l)$ plane 
      for $p_0=\pm p$.  The area shaded in dark gray is excluded by
      the delta functions. The region shaded in light gray is below
      the light-cone (dotted lines).  The thin solid curves are the
      transverse dispersion curves of the thermalized gluon. The
      vertical dotted line is the separation between
      $\epsilon(p_0)\epsilon(r_0+l_0)=+1$ and
      $\epsilon(p_0)\epsilon(r_0+l_0)=-1$.}}
  \label{fig:kine}
\end{figure}

\begin{figure}[htbp]
  \centerline{
    \resizebox*{!}{4cm}{\includegraphics{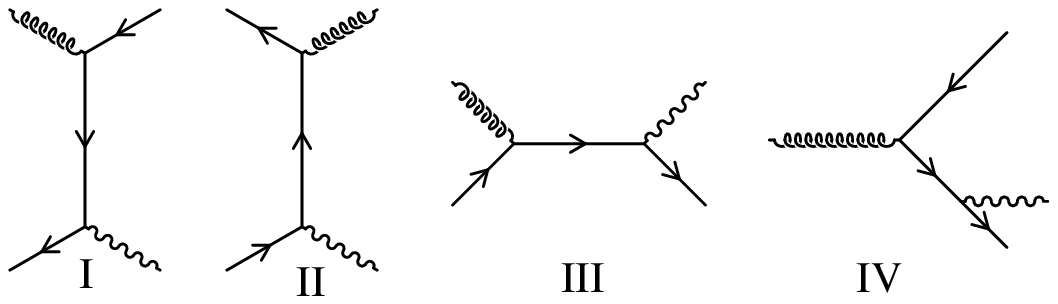}}
    }
  \caption{\footnotesize{Examples of physical processes 
      included in the phase-space of of Fig.~\ref{fig:kine}.  Region
      I: $p_0<0$, $r_0+l_0<0$: Compton scattering of an antiquark.
      Region II: $p_0<0$, $r_0+l_0>0$: $q\bar{q}$ annihilation.
      Region III: $p_0>0$, $r_0+l_0>0$: Compton scattering of a quark.
      Region IV: $p_0>0$, $r_0+l_0<0$: plasmon decay into $q\bar{q}$
      and a photon. For the gluon, we use an approximate dispersion
      relation, with a constant thermal mass. The quarks being hard
      need not be effective.}}
  \label{fig:physpro1}
\end{figure}

Being dominated by hard $p$, we can neglect $m_{_{F}}$ in ${\rm
  Im}\,\Pi_{1}$, and hence we can use the same kinematical
approximations as in \cite{AurenGKZ1}.  Using the $\delta((R+L)^2)$
function, we get for the angle $\theta^{\prime}$ between ${\imb p}$
and ${\imb l}$:
\begin{equation}
  \cos\theta^\prime={{(r_0+l_0)^2-p^2-l^2}\over{2pl}}\; .
  \label{eq:cosine}
\end{equation}
Additionally, $\cos\theta^\prime$ must be kept within $[-1,+1]$, which
places some constraints on the phase space. Eq.~(\ref{eq:cosine}),
together with the condition on $\cos\theta^\prime$, gives the following
inequalities:
\begin{eqnarray}
  &&(l_0-l+p_0+q_0-p)(l_0+l+p_0+q_0+p)\le 0 \\
  &&(l_0-l+p_0+q_0+p)(l_0+l+p_0+q_0-p)\ge 0\; ,
\label{eq:ineq}
\end{eqnarray} with $|p_0| = p$. We assume $p$ fixed and hard and solve
the inequalities for $l_0$ and $l$ which leads to the phase space
reduction seen in Fig.~\ref{fig:kine}. 
The different allowed regions in Fig.~\ref{fig:kine} can be
interpreted in terms of physical processes (Fig.~\ref{fig:physpro1})
determined by the relative signs of the energies $p_0$, $l_0$ and
$r_0+l_0$. We have Compton processes in regions (I) and (III),
quark-antiquark annihilation in region (II), and plasmon decay in
region (IV). The symmetry of the integrand in ${\rm Im}\,\Pi_{1}$ with
respect to the change of variables $P\leftrightarrow -R-L$, indicates
that regions (I) and (III) give equal contributions. Therefore, we
will consider one region (I for instance) and multiply the end result
by $2$.  The domain of integration of Eq.~(\ref{eq:formula}) is
constrained by the $\delta$ function Eq.~(\ref{realpi}), which further
reduces the allowed phase space to the intersection of the dispersion
curves with the regions I to IV. We also note that the
longitudinal modes of the gluon cannot give a large logarithm in the
momentum integral, since longitudinal modes are exponentially
suppressed in the hard region.

The transverse gluon polarization tensor introduced in
Eq.~(\ref{realpi}) is given by:
\begin{equation}    
\Pi_{_{T}}(l_0,l) = 3m_{\rm g}^2\left[
    {{x^2}\over 2}+{{x(1-x^2)}\over{4}}\ln\left|
      {{x+1}\over{x-1}}\right|\right]\; ,
\label{eq:gluon}
\end{equation}
where $x\equiv l_0/l$.  For a time-like gluon, $\Pi_{_{T}}(lx,l)$ is a
function of $x$ varying slowly between $m_{\rm g}^2/2$ and $3m_{\rm
  g}^2/2$.  Therefore, at the logarithmic accuracy in the evaluation
of Eq.~(\ref{eq:formula}), we can take $\Pi_{_{T}} \approx m_{\rm
  g}^2$.  Making this approximation, we obtain the following
expression for ${\rm Im}\,\Pi_{1}$ :
\begin{eqnarray}
&&{\rm Im}\,\Pi_{1}(q_0,{\imb 0)}\approx 
  -{{8NC_{_{F}}e^2g^2}\over{(2\pi)^3}} {q_0}
  \int\limits_{0}^{+\infty}dp\,
  n^\prime_{_{F}}(p)\;
  \int\limits_{-\infty}^{-1}dx\nonumber\\
 &&\qquad\times\int\limits_{{q_0}\over{1-x}}^{-{q_0}\over{1+x}}
  dl\,l^2\delta(L^2-m_{\rm g}^2)[n_{_{B}}(l x)+n_{_{F}}(-p+q_0+l x)]
\end{eqnarray}
where we have used $n_{_{F}}(p_0+q_0) - n_{_{F}}(p_0) \approx q_0
n^\prime_{_{F}}(p)$. In the normalization, we have taken into account
the above mentioned factors of $2$. The integrals can easily be done to
logarithmic accuracy and we obtain:
\begin{equation}
 {\rm Im}\;\Pi_{1}(q_0,{\imb 0})\approx - {{NC_{_{F}}e^2g^2}\over{4\pi^3}}
\Big(\sum_f e^2_f\Big){q_0}T 
 \ln\left(\frac{q_0T}{m_{\rm g}^2+ q_0^2}\right)\; ,
\label{eq:hardp2l}
\end{equation}
where we have reintroduced the summation over the flavors running in
the quark loop.  The occurrence of $m_{\rm g}$ indicates that this
logarithm cannot be found at one-loop in the HTL scheme since only the
thermal fermion mass appears at this level. As we will see later on,
this is not compensated by the counterterms. We find that the
annihilation process (region II) contributes to ${\rm Im}\;\Pi_{1}$ at
leading order $e^2g^2 q_0 T$, but without a logarithm since $l$ is
constrained to be hard in this region (see Fig.~\ref{fig:kine}).
Similarly, the plasmon decay process (region IV) can be ignored at the
logarithmic accuracy.

\subsection{Hard $l$ region (${\rm Im}\,\Pi_{2}$)}
In order to calculate ${\rm Im}\,\Pi_{2}$, we follow the same
procedure as in the previous section but interchanging the roles of
$l$ and $p$. Since ${\rm Im}\,\Pi_{2}$ is dominated by the hard
$l$-region, we can neglect the gluon thermal mass. Again we start with
the $\delta$ function $\delta((R+L)^2-m_{_{F}}^2)$, which provides
us with the angle $\theta^{\prime}$ via the relation:
\begin{equation}
  \cos\theta^\prime={{R^2-m^2_{_{F}}+2r_0l_0+L^2}\over{2rl}}\; .
\end{equation}

\begin{figure}[hbt]
  \centerline{\resizebox*{!}{5cm}{\includegraphics{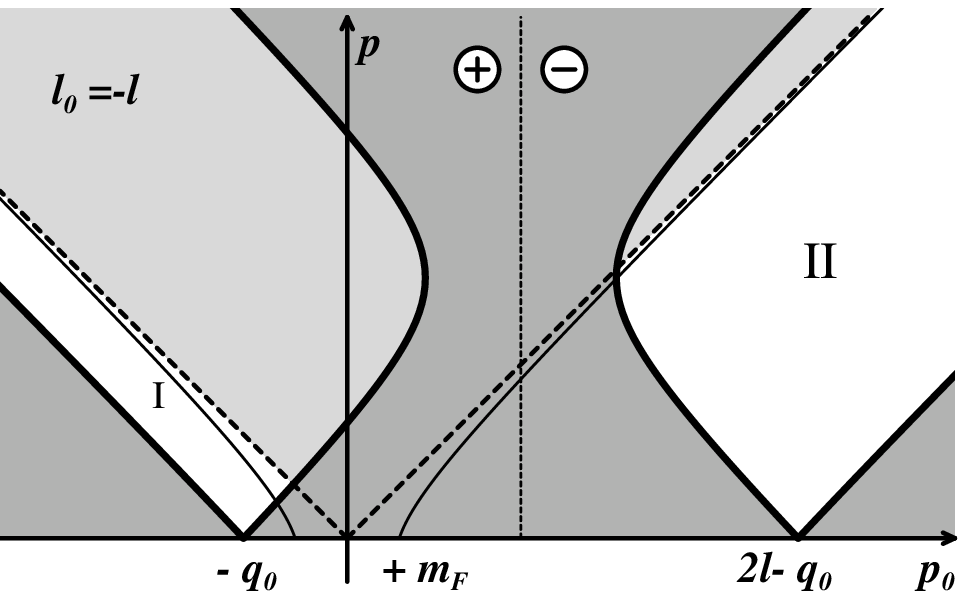}}}
  \centerline{\resizebox*{!}{5cm}{\includegraphics{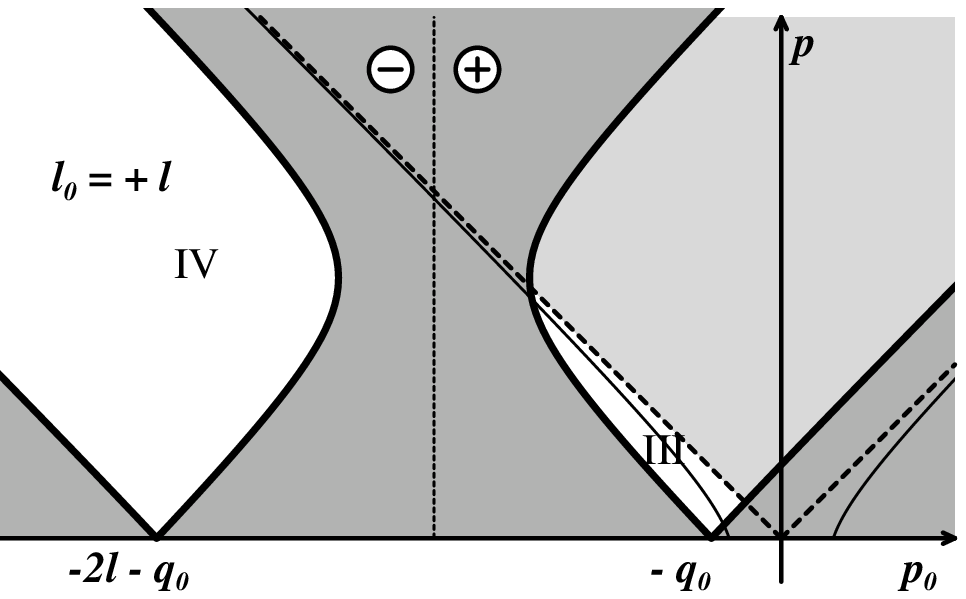}}}
  \caption{\footnotesize{Allowed domains in the $(p_0,p)$ plane for 
      $l_0=\pm l$.  The area shaded in dark gray is excluded by the
      delta function constraints. The areas shaded in light gray are
      below the light-cone (dotted lines). The thin solid curves are
      the mass shells of the thermalized quarks.  The vertical dotted
      line is the separation between
      $\epsilon(l_0)\epsilon(r_0+l_0)=+1$ and
      $\epsilon(l_0)\epsilon(r_0+l_0)=-1$.}}
  \label{fig:kinlhard}
\end{figure}
As before, we must enforce $-1\le\cos\theta^\prime\le 1$, which
implies the following set of inequalities:
\begin{eqnarray}
  &&(r_0-r+l_0+l)(r_0+r+l_0-l)\ge m^2_{_{F}}\\
  &&(r_0-r+l_0-l)(r_0+r+l_0+l)\le m^2_{_{F}}\; .
  \label{hardconstraints}
\end{eqnarray}
\noindent We can further simplify these inequalities and write them as:
\begin{eqnarray}
 &&\sqrt{(p-l)^2+m_{_{F}}^2}\le l_0+r_0\ {\rm or}\ 
  l_0+r_0 \le -\sqrt{(p-l)^2+m_{_{F}}^2}\\
&& -\sqrt{(p+l)^2+m_{_{F}}^2}\le l_0+r_0 \le \sqrt{(p+l)^2+m_{_{F}}^2}\; .
\label{eq:hardlreg}
\end{eqnarray}
Keeping $l$ hard, the above inequalities lead to a reduction of the
allowed domain in the $(p_0,p)$ plane (see Fig.~\ref{fig:kinlhard}) where
the case $l_0=l$ and $l_0=-l$ have to be distinguished.

As before, the various regions admit a physical interpretation.
Regions I and II represent Compton scattering on an antiquark and on a
quark respectively, while region III contains quark-antiquark
annihilation.  Region IV describes quark decay into quark, gluon and
photon, but it is not allowed if the initial and final quarks have the
same masses. We note also the absence of plasmon decay, which is
entirely due to the different approximations made in this section and
in the previous one: here we assume $L^2=0$ but keep the fermion mass
and this forbids the decay of the gluon into a photon and a massive
quark-antiquark pair.

\begin{figure}[htbp]
  \centerline{
    \resizebox*{!}{4cm}{\includegraphics{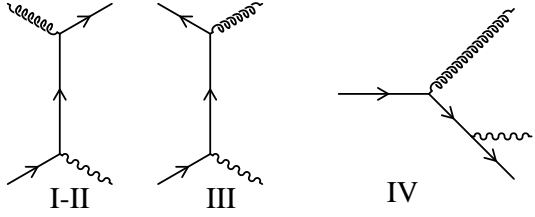}}
    }
  \caption{\footnotesize{Examples of physical processes 
      included in the phase-space of Fig.~\ref{fig:kinlhard}.  Region
      I: $p_0<0$, $r_0+l_0<0$,$l_0<0$: Compton scattering of an
      antiquark.  Region II: $p_0>0$, $r_0+l_0>0$,$l_0<0$: Compton
      scattering of a quark.  Region III: $p_0<0$,
      $r_0+l_0>0$,$l_0>0$: $q\bar{q}$ annihilation.  Region IV:
      $p_0<0$, $r_0+l_0<0$,$l_0>0$: quark decay into $qg$ and a
      photon. For the quarks, we use an approximate dispersion
      relation, with a constant thermal mass. The gluon being hard is
      treated as a bare gluon.}}
  \label{fig:phyproc2}
\end{figure}
The change of variables $P\leftrightarrow -R-L$ in the $Q\cdot L /
(\overline {P+L})^2$ term in ${\rm Im}\,\Pi_{2}$ allows us to
write:
\begin{eqnarray}
&&
  \!\!\!{\rm Im}\,\Pi _{2}(q_0,{\imb q}) \approx 
   - 8 NC_{_{F}} e^2g^2 
   \int{{d^4P}\over{(2\pi)^3}}\int{{d^4L}\over{(2\pi)^3}}
  [n_{_{F}}(r_{0})-n_{_{F}}(p_{0})]\nonumber\\
  &&\times\;
  [n_{_{B}}(l_{0})+n_{_{F}}(r_{0}+l_{0})]\delta(\overline{P}^{2})
  \delta((\overline{R+L})^2)\epsilon(p_{0})\epsilon(r_{0}+l_{0})\nonumber\\
  &&\times\;
  2\pi \epsilon(l_{0})\delta(L^2)\frac{Q\cdot L}{\overline{R}{}^{2}}
\label{eq:imtwo}
\end{eqnarray} 
This integral can be evaluated relatively easily. Consider, for
example, region III ($l_0=l, \ p_0 < 0, \ r_0+l_0 > 0, \ i.e. \ q \bar
q$ annihilation). For $l$ fixed and hard, the fermion dispersion curve
$p_0 \approx -\omega_p = \sqrt {p^2 + m_{_F}^2}$ intersects the
boundary of the region at
\begin{eqnarray}
p_{\rm min} \approx {|q_0^2 - m_{_F}^2| \over 2 q_0}, \qquad
p_{\rm max} \approx l 
\end{eqnarray}
so that, after doing all angular integrations and using the
$\delta$ functions, one finds
\begin{eqnarray}
{\rm Im}\,\Pi^{^{III}}_{2} &\approx& - NC_{_{F}} \frac{e^2g^2}{(2\pi)^3} q_0
\int\limits_{0}^{\infty} ldl\,  \int\limits_{p_{\rm min}}^{p_{\rm max}} 
{p dp \over \omega_p} \nonumber \\
&&\quad \times[n_{_{B}}(l)+n_{_{F}}(l-\omega_p)] 
\ {[n_{_{F}}(\omega_p)-n_{_{F}}(\omega_p - q_0)] \over 2 q_0 \omega_p - q_0^2}
 \nonumber \\
&\approx& - {NC_{_{F}} \over 2} \frac{e^2g^2}{(2\pi)^3}
\int\limits_{0}^{\infty} ldl\ \int\limits_{|q_0^2 + m_{_F}^2| \over 2
q_0}^{l} {d\omega_p} \nonumber \\
&& \quad \times[n_{_{B}}(l)+n_{_{F}}(l-\omega_p)] 
\ {[n_{_{F}}(\omega_p)-n_{_{F}}(\omega_p - q_0)] \over \omega_p - q_0/2}
\label{hardl}
\end{eqnarray}
We can neglect $\omega_p$ in front of $l$ in the term
$n_{_{F}}(l-\omega_p)$, and the two integrals decouple. The integral
over $l$ leads to the usual hard thermal factor ${\pi^2T^2}/{4}$ while
the integral over $p$ yields a logarithmic factor $\ln( {q_0 T /
  m_{_F}^2} )$.  Using the same method, one finds that the
contribution of the sector $l_0=-l$ in Fig.~\ref{fig:kinlhard} is
similar to Eq.~(\ref{hardl}) so that, to the required accuracy,
Eq.~(\ref{eq:imtwo}) leads to
\begin{equation}
 {\rm Im}\,\Pi_{2}(q_0,{\imb 0})         
\approx - \frac{NC_{_{F}}}{8} \frac{e^2g^2}{4\pi}
\Big(\sum_f e_{f}^2\Big) q_0 T \ln\left( {q_0 T \over m_{_F}^2} \right)
\label{eq:hardl2l}
\end{equation}
where we have included the charge factor of the quarks. The
normalization factor takes into account the contribution of all the ways
of cutting through the photon polarization diagrams. 


\section{Counterterms}

We have calculated the two-loop diagrams using various approximations.
As a consequence, in order to be consistent, we have to use the same
approximation when evaluating the counterterms. In the appendix, we
show that the complete computation of the counterterms gives the same
logarithmic contribution as the simplified version of the counterterm
diagrams we discuss in this section.


The simplified version of the vertex counterterm and the self-energy
counterterm are depicted on the right of Fig.~\ref{fig:2loopsm}.   

Their general structure is as in Eq.~(\ref{2loop}) but the integrals
have now to be evaluated using the HTL approximations, namely $L$ hard
with $L^2=0$ and $P$ and $R$ neglected with respect to $L$ with, as
before, the notation
$\overline{P}=(p_{0},\sqrt{p^2+m_{_{F}}^2}\hat{\imb p})$,
$\overline{R}=(p_{0}+q_{0},\sqrt{p^2+m_{_{F}}^2}\hat{\imb
  p})$. This implies $(\overline{R+L})^2 \approx 2 {R} \cdot L$.  One
then writes
\begin{eqnarray}
&& {\rm Im}\,\Pi(q_0,{\imb 0})|_{\rm ct} \approx
\frac{NC_{_{F}}}{2} e^{{2}}g^{{2}}
\int{{d^4P}\over{(2\pi)^3}}\int{{d^4L}\over{(2\pi)^3}}
[n_{_{F}}(r_{0})-n_{_{F}}(p_{0})] \nonumber \\
&& \qquad \qquad \qquad \qquad 
\times \; [n_{_{B}}(l_{0})+n_{_{F}}(l_{0})]
\epsilon(p_{0})\delta(\overline{P}^{2})\delta(2R\cdot L)\delta(L^2)
 \nonumber \\
&& \qquad \qquad \qquad \qquad 
\times \; 2 \pi g_{\rho\sigma}\left[  {{{\rm Trace^{\rho\sigma}_{_{HTL}}}{|_{\rm vertex}}} 
\over{(2P\cdot L)\overline{R}{}^2}} 
+{{{\rm Trace^{\rho\sigma}_{_{HTL}}}{|_{\rm self}}} 
\over{\overline{R}{}^2 \overline{R}{}^2}}\right]
\label{eq:counterm}
\end{eqnarray}
where the traces can be evaluated using the Feynman gauge for the gluon.
In this equation we make use of $r_0+l_0 \approx l_0$ so that some
$\epsilon$-functions drop out. 

For the vertex we find simply in the HTL approximation
\begin{equation}
   g_{\rho\sigma}{\rm Trace^{\rho\sigma}_{_{HTL}}}|_{\rm vertex} \approx 32\ 
\overline{R}.L \ \overline{P}.L=0\; .
\end{equation}
\noindent 
Hence the vertex counterterm is vanishing within the approximation
used. This is not a general result but a consequence of
using the Feynman gauge.


The self energy counterterm is non vanishing and we find
\begin{eqnarray}
g_{\rho\sigma}{\rm Trace^{\rho\sigma}_{_{HTL}}}|_{\rm self} &=& 16\ 
( 2 (\overline {P}\cdot\overline {R})
(\overline {R} \cdot L) - \overline R^2 \overline {P} \cdot L)\nonumber \\
 &\approx& 16 \ \overline R^2  \ Q\cdot L \ .
\label{eq:selfct}
 \end{eqnarray}
 In the last line, we have neglected terms which are suppressed by
 inverse powers of $p$ at large $p$ which cannot contribute to
 a logarithmic factor in the $p$ integration.  Plugging this
 expression in Eq.~(\ref{eq:counterm}) and comparing to
 Eq.~(\ref{eq:imtwo}), we find that these two equations coincide at
 the logarithmic level, except for the sign. Indeed, their sum
 receives a contribution only when $p$ is hard and therefore does not
 have a logarithmic behavior. One concludes that, to logarithmic
 accuracy,
\begin{equation}
{\rm Im}\,\Pi_{2}(q_0,{\imb 0}) +
{\rm Im}\,\Pi(q_0,{\imb 0})|_{\rm ct} \simeq 0
\label{eq:CT}
\end{equation}
so that, as expected, the counterterms cancel the two-loop
contribution arising from the hard $L$ phase-space.

\section{The total two-loop contribution}
\label{sec:tot2loop}

We summarize in this section the complete results of the calculation
of the virtual photon rate up to two loops in the effective
perturbative expansion.  We have to collect the following pieces: the
one-loop result of \cite{BraatPY1} given in Eq.~\ref{eq:1loop}, the
two-loop bremsstrahlung contribution derived in \cite{AurenGKZ1} and
finally the contribution of the Compton and annihilation processes we
have calculated in the previous sections of the present paper. Adding
all three contributions, we obtain to leading order in
$\ln(1/g)$:
\begin{eqnarray}
        \frac{dN}{d^4xdq_0 d^3{\imb q}}&\approx&
        \frac{\alpha^2}{3\pi^6 q_0^2}\Big(\sum_{f}e^2_f\Big)
        \left(\frac{NC_{_{F}}g^2T^2}{8}\right)\nonumber\\
&&\times\left\{ \frac{\pi^2 m_{_{F}}^2}{4q_0^2}\ln\left( {{T^2}
      \over{m^2_{_{F}}}} \right)
       +\frac{3m^2_{\rm g}}{q_0^2}\ln\left(\frac{T^2}{m_{\rm g}^2}\right)
\right.\nonumber\\     
&& \left.+ \frac{\pi^2}{4}
      \ln\left(\frac{q_0T}{m^2_{_{F}}+q_0^2}\right)
     +2\ln\left(\frac{q_0 T}{m_{\rm g}^2+ q_0^2} \right) \right\}\; .
\label{eq:tottwo}
\end{eqnarray}
The term in ${m_{_{F}}^2}/{q_0^2}$ arises from the cut-cut term in the
one-loop calculation while the term in ${m^2_{\rm g}}/{q_0^2}$
describes the bremsstrahlung coming from the two-loop diagrams. The
last line combines the pole-cut contribution at one-loop,
Eq.~(\ref{eq:1loop}), as well as the $L^2> 0$ pieces at two-loop,
Eqs.~(\ref{eq:hardp2l}), (\ref{eq:hardl2l}), and the associated
counterterms Eq.~(\ref{eq:CT}). In the above equation, the choice of
hiding some powers of $gT$ in thermal masses is somewhat arbitrary.
Nevertheless, it should be noted that, even if the four terms have the
same order of magnitude when $q_0$ is soft, two of them are
proportional to $e^2g^4$ and therefore would be obtained only at three
loops in the naive perturbative expansion. 

If we consider the leading term in an expansion in powers of $g$ at
fixed $q_0$ (which is precisely what one is doing in the bare theory),
we should be able to recover the results of two-loop bare calculations
performed by \cite{AltheA1}. Taking this limit in
Eq.~(\ref{eq:tottwo}), we find:
\begin{equation}
  \frac{dN}{d^4xdq_0 d^3{\imb q}}\approx
        \frac{\alpha^2}{3\pi^6 q_0^2}\Big(\sum_{f}e^2_f\Big)
        \left(\frac{NC_{_{F}}g^2T^2}{8}\right)
        (\frac{\pi^2}{4}+2)\ln({T\over q_0})\; .
\label{eq:bare}
\end{equation}
\noindent
which is precisely the result found in \cite{AltheA1}.  One may notice
that, had we kept only the terms calculated by BPY~\cite{BraatPY1}, the
$g\to 0$ expansion of Eq.~(\ref{eq:tottwo}) would not
reproduce the complete bare result.

One may also wonder what happens when considering the ultra-soft
photon limit, $q_0 \ll m_{_F}$ or $q_0 \ll m_{\rm g}$, for which the
pole-cut contributions in Eq.~(\ref{eq:tottwo}) seems completely
wrong.  A glance at Fig.~\ref{fig:kine} and Fig.~\ref{fig:kinlhard}
shows that, in that case, the intersection of the dispersion curves
with the boundaries of the physical regions are pushed to hard $l$ or
$p$ values so that no logarithm is generated.  This is reflected in
Eq.~(\ref{eq:tottwo}), where some logarithms become small and
eventually negative when $q_0$ goes to zero, indicating that the terms
we considered no longer exhibit a large logarithmic factor.  More
technically, in the case of Eq.~(\ref{eq:hardp2l}) for example, the
integration generates terms such as $$
\ln(1-\exp(-\frac{q_0^2+m_{\rm
    g}^2}{2 q_0 T}))$$
which reduce to the usual logarithm for the
generic case $q_0\sim m_{\rm g}\sim g T$ but which lead to an
exponentially suppressed factor when $q_0 \ll m_{\rm g}^2/T$.

For completeness, we briefly mention here the case of thermal
production of hard or very hard real photon ($q_0 \ge
T$)~\cite{BaierNNR1,KapusLS1}. It was found in the 2-loop
approximation~\cite{AurenGKZ1} that, as in the case of soft virtual
photon production, leading contributions arose from the space-like
part of the gluon spectral density ($L^2 < 0$) associated to the
bremsstrahlung process as well as to $q-\bar q$ annihilation with
scattering. To complete the calculation, the time-like ($L^2 \ge 0$)
part should be considered. However it is not necessary to do the
calculation using the counterterm method advocated above since in the
published works~\cite{{BaierNNR1},{KapusLS1}} the two-loop
contribution has already been included using the cutoff method.  The
final result contains a factor $\ln(q_0 T/ m_{_F}^2)$ similar to those
of Eq.~(\ref{eq:tottwo}).

\section{Conclusions}
In this work, we complete the study of soft static lepton pair
production in a quark-gluon plasma, in the two-loop approximation of
the effective perturbative expansion. At one loop, the result shows a
logarithmic sensitivity to hard momenta in the loop. Since the
extrapolation of hard thermal loops is not accurate enough in the hard
space-like region, we anticipate that the one-loop approximation may
not yield the complete result at the leading order in the expansion in
powers of the coupling constant. We should then consider two-loop
diagrams, and an explicit calculation shows that, indeed, they give a
contribution at leading order. Two types of terms can be
distinguished, which are associated with different physical processes:

\noindent(i) two-loop leading corrections to processes 
already included at one-loop. They are the Compton and annihilation
production of the virtual photon, which are calculated 
incompletely at the one-loop level.

\noindent(ii) New processes, not contained in the one-loop 
approximation, namely the bremsstrahlung production of the photon.
This was studied elsewhere \cite{AurenGKZ1}.

Dealing with two-loop diagrams in the HTL resummed theory requires
some care. Two approaches are possible. One can use the cutoff
method where the two-loop corrections are taken into account only
above some cutoff value of the loop momentum. But this is not enough to
evaluate the two-loop diagrams using bare propagators and vertices
only. On the contrary, effective propagators and vertices are
necessary when appropriate. Failing to do this, one could miss the
``new'' contributions referred to above.

As an alternative to the cutoff method,
 one can construct the perturbative expansion from the effective
Lagrangian, and counterterms must then be included to avoid double
counting. These are crucial to correctly calculate the physical
processes which appear at different orders of the loop expansion of
the effective theory. This is the approach advocated in this work.

Since the new production mechanism of virtual photon at two-loop order
shows a logarithmic sensitivity to hard space-like momenta in the
loop, it cannot be claimed that our result is complete, for the same
reason that the one-loop prediction could not be trusted. The
contribution of three-loop diagrams should be considered. However, since
in the present calculation there is no physical process that would
arise at four loops or more in the bare theory, leading contributions
coming from more than three loops are not expected.

\appendix
\section{Counterterms}
\subsection{Preliminaries}
In this appendix we give the exact evaluation of the counterterms. We
show that the logarithmic behavior coincides with those found in the
paper using the simplified version of the counterterm diagrams.  The
result according to which the HTL part of the effective vertices does
not modify the calculation of the logarithmic part is to be expected
on the basis of a quite general argument. Indeed, we know that the HTL
correction to the $q\bar{q}\gamma$ vertex behaves like:
\begin{equation}
e(gT)^2\int{d\Omega_{\imb l}}{{\hat{\slL}\hat{L}^\mu}\over{(P\cdot\hat{L})
(R\cdot\hat{L})}}\; ,
\end{equation}
where $L\equiv(1,\hat{\imb l})$ and where $P$ and $R$ are the momenta
of the quark and of the antiquark. This means that this HTL correction
has a very different scaling behavior when $P$ and $R$ become hard,
compared to the bare part of the effective vertex (which is
independent of $P$ and $R$). As a consequence, we do not expect a
contribution to the logarithmic part from the HTL correction to the
vertices. This is what we check explicitly in the following.

To calculate exactly the counterterms of Fig.~\ref{fig:2loop}, we
follow \cite{BraatPY1}, where the authors relate the imaginary part of
the effective vertex to that of the effective propagator, in order to
write the final result in terms of the propagator spectral density
only. In this paragraph, we define some notations, and some useful
formulae.

The effective propagator can be written as:
\begin{eqnarray}
   ^{*}{\cal S}(p_0,{\imb p})&=& \sum_{\tau=\pm 1}\frac{\hat{\slP}_{\tau}}
   {2D_{\tau}(p,p_0/p)}\nonumber\\
   {\rm with}\qquad D_{\tau}(p,x)&\equiv& \frac{1}{p}[p^2(x-\tau)-m_{_{F}}^2\
   tau+m_{_{F}}^2
   (\tau x-1)Q_0(x)]\; ,
\label{eq:propD}
\end{eqnarray}
where
$\hat{\slP}_\tau\equiv\gamma^o-\tau\mbox{\boldmath$\gamma$}\cdot{\hat{\imb
    p}}$, and $Q_0(x)\equiv \frac{1}{2}\ln(({x+1})/({x-1}))$ is the first
Legendre function.  The presence of Legendre functions in the vertex
and the propagator allows us to relate the imaginary part of the
vertex (or the imaginary part of the product of a propagator with a
vertex) to the imaginary part of the propagator.  The spectral density
of a quark and the imaginary part of the product of a vertex with a
propagator are given by:
\begin{eqnarray} 
    &&\!\!\!\!\!\rho_{\tau}(p,x)
=-2{\rm Im}\frac{1}{D_{\tau}(p,x)}
=\frac{p^2(x^2-1)}{2m_{_{F}}^2}[\delta(px-\omega_{\pm}(p))
+\delta(px+\omega_{\mp}(p))]\nonumber\\
  && \qquad\qquad\qquad\qquad\qquad\qquad +\beta_{\pm}(p,x)\theta(1-x^2)\; ,\\
&&{\rm Im}\left(\frac{Q_0(x)}{D_{\tau}(p,x)}\right)=\frac{1}{2}\left(
  \frac{p^2(x-\tau)-m_{_{F}}^2\tau}{m_{_{F}}^2(\tau x-1)}\right)\rho_{\tau}
 (p,x)
\label{eq:spectfer}
\end{eqnarray}
with
\begin{equation}
 \beta_{\pm}(p,x)=\frac{(m_{_{F}}^2/2p^3)(1\mp x)}{\left(x\mp 1-(m_{_{F}}/p)^2 
   [Q_0(x)\mp Q_{1}(x)]\right)^2+[(\pi m_{_{F}}^2/2p^2)(1\mp x)]^2}
\end{equation} 
where $Q_{1}(x)\equiv xQ_0(x)-1$ is the second Legendre function, and
$\omega_{\pm}(p)$ are the solution of the dispersion equation
$D_{\tau}(p,\omega_\pm(p)/p)=0$.

We need also a formula giving the imaginary part of the product of two
vertices and a propagator, given by:
\begin{eqnarray}
{\rm Im}\;\left(\frac{Q_0^2(x)}{D_{\tau}(p,x)}\right)=\frac{p\;{\rm
Im}Q_0(x)}{m_{_{F}}^2(\tau x-1)}
-\frac{1}{2}\left(
\frac{p^2(x-\tau)-m_{_{F}}^2\tau}{m_{_{F}}^2(\tau x-1)}\right)^2
\rho_{\tau}(p,x)  
\end{eqnarray}
where in  the retarded prescription ($x\rightarrow x+i\epsilon$):
\begin{equation}
 {\rm Im}Q_0(x)=-\frac{\pi}{2}\theta(1-x^2)\; .
\end{equation}

\subsection{Vertex counterterms}
There are two counterterms diagrams associated with the 2-loop vertex
diagram: one is shown in the lower right corner of
Fig.~\ref{fig:2loop} and the other one with the counterterm in place
of the other loop. In what follows, we denote by $-R$ the momenta
circulating in the upper quark of the diagram of Fig.~\ref{fig:2loop}
and $P$ to be the momenta of the lower quark.

Applying the previous formulae, and following \cite{AurenBP1}, we get:
\begin{eqnarray}  
  &&\left.\frac{dN}{d^4xdq_0 d^3{\imb q}}\right|_{\rm vertex}
=\frac{4N(\sum\nolimits_f e_{f}^2)}
  {3\pi^4}\frac{\alpha^2}{q_0^2}\int\limits_{0}^{+\infty}p^2dp
\int\limits_{-\infty}^{+\infty}dp_0
\int\limits_{-\infty}^{+\infty}dr_0\;n_{_{F}}(p_0)n_{_{F}}(r_0)\nonumber\\
  &&\times\;
  \delta(q_0-p_0-r_0)\left\{ 2\left(1-\frac{p_0^2-r_0^2}{2q_0 p}\right)^2
\rho_{+}(P)\rho_{-}(R)\right.\nonumber\\
&&
-\left(1-\frac{p_0^2-r_0^2}{2q_0 p}\right)\rho_{+}(P)\rho_{-}(R)\nonumber\\
  &&+ 2\left(1+\frac{p_0^2-r_0^2}{2q_0p}\right)^2\rho_{-}(P)
  \rho_{+}(R)-\left(1-\frac{p_0^2-r_0^2}{2q_0 p}\right)\rho_{-}(P)
  \rho_{+}(R)\nonumber\\
  &&+\left(1+\frac{p_0^2+r_0^2-2p^2-2m_{_{F}}^2}{2q_0 p}\right)^2\rho_{+}(P)
  \rho_{+}(R)\nonumber\\
  &&-\left(1+\frac{p_0^2+r_0^2-2p^2-2m_{_{F}}^2}{2q_0p}\right)
  \rho_{+}(P)\rho_{+}(R)\nonumber\\
  &&+\left(1-\frac{p_0^2+r_0^2-2p^2-2m_{_{F}}^2}{2q_0 p}\right)^2\rho_{-}(P)
  \rho_{-}(R)\nonumber\\
  &&-\left(1-\frac{p_0^2+r_0^2-2p^2-2m_{_{F}}^2}{2q_0 p}\right)
  \rho_{-}(P)\rho_{-}(R)\nonumber\\
  &&+\theta(p^2-p_0^2)\frac{m_{_{F}}^2}{8pq_0^2}(1-x^2)[(1+x)\rho_{+}(R)+(1-x)
  \rho_{-}(R)]\nonumber\\
  &&\left.+\theta(p^2-r_0^2)\frac{m_{_{F}}^2}{8pq_0^2}(1-\left(\frac{r_0}{p}\right)^2)
  [(1+\frac{r_0}{p})\rho_{+}(P)+(1-\frac{r_0}{p})\rho_{-}(P)]\right\}
\label{eq:countvert}
\end{eqnarray}

Since we want to look for possible double counting in Compton and
annihilation processes, we must keep at least one time-like quark
momentum ($R^2\ge 0$ or $P^2\ge 0$). By symmetry, we can limit
ourselves to one of these two regions, and multiply the end result by
a factor of $2$.  Doing the logarithmic approximation ({\it i.e} taking
$\beta_{\pm}(p,x)\approx {m_{_{F}}^2}/{2p^3(1\mp x)}$, ....), we
deduce that the vertex counterterm vanishes at the level of
approximation:
\begin{equation} 
    \left.\frac{dN}{d^4xdq_0 d^3{\imb q}}\right|_{\rm vertex}
    \build{\approx}\over{{\rm log}} 0\; ,
\end{equation}
which is equivalent to the result found using the simplified version
of the counterterm diagram.

\subsection{Propagator counterterms}
Making use of the same tools, we can calculate the counterterm diagram
associated with the self-energy correction (diagram in the upper right
corner of Fig.~\ref{fig:2loop}, with the other symmetric diagram where
the counterterm insertion is on the lower quark propagator).

Now, the relevant region of phase-space that we want to consider is
the region where the line having a counterterm insertion is space-like
while the other one is time-like.  This is a consequence of limiting
ourselves to the cut passing through the self energy insertion in the
two loops diagram.

After a lengthy but direct calculation we get (including the two
possible diagrams of counterterms insertion):

\begin{eqnarray}  
  &&\left.\frac{dN}{d^4xdq_0 d^3{\imb q}}\right|_{\rm self}
=\frac{4N(\sum\nolimits_f e_{f}^2)}{3\pi^4}
\frac{\alpha^2}{q_0^2}
\int\limits_{0}^{+\infty}p^2dp\int\limits_{-\infty}^{+\infty}dp_0
\int\limits_{-\infty}^{+\infty}dr_0\;n_{_{F}}(p_0)n_{_{F}}(r_0)\nonumber\\
  &&\times\;
  \delta(q_0-p_0-r_0)\theta(p^2-p_0^2)\nonumber\\
&&\times
  \left\{ 2\left(1-\frac{p_0^2-r_0^2}{2q_0 p}\right)^2\left[-\frac{p^2}{2}
  \frac{\partial}{\partial
  p}\left(\frac{\beta_{+}(P)}{p}\right)-\beta_{+}(P)\right]\Delta_{-}(R)\right.\nonumber\\
  &&+2\left(1+\frac{p_0^2-r_0^2}{2q_0 p}\right)\left[-\frac{p^2}{2}
  \frac{\partial}{\partial  p}\left(\frac{\beta_{-}(P)}{p}\right)-\beta_{-}(P)
  \right]\Delta_{+}(R)\nonumber\\
  &&+\left(1+\frac{p_0^2+r_0^2-2p^2-2m_{_{F}}^2}{2q_0 p}\right)^2
  \left[-\frac{p^2}{2}
  \frac{\partial}{\partial  p}\left(\frac{\beta_{+}(P)}{p}\right)-\beta_{+}(P)
  \right]\Delta_{+}(R)\nonumber\\
  &&+\left(1-\frac{p_0^2+r_0^2-2p^2-2m_{_{F}}^2}{2q_0 p}\right)^2
  \left[-\frac{p^2}{2}
  \frac{\partial}{\partial  p}\left(\frac{\beta_{-}(P)}{p}\right)-\beta_{-}(P)
  \right]\Delta_{-}(R)\nonumber\\
  &&-\frac{m_{_{F}}^2}{8pq_0^2}(1-x^2)[(1+x)\Delta_{+}(R)+(1-x)\Delta_{-}(R)]
  \nonumber\\
  &&+2\frac{p}{q_0}x(x-1)\left(1-\frac{p_0^2-r_0^2}{2q_0p}\right)\rho_{+}(P)
  \Delta_{-}(R)\nonumber\\
  &&-2\frac{p}{q_0}x(x+1)\left(1+\frac{p_0^2-r_0^2}{2q_0p}\right)\rho_{+}(P)
  \Delta_{+}(R)\nonumber\\
  &&-\frac{p}{q_0}(x^2-1)\left(1+\frac{p_0^2+r_0^2-2p^2-2m_{_{F}}^2}{2pq_0}\right)
  \beta_{+}(P)\Delta_{+}(R)\nonumber\\
  &&\left.+\frac{p}{q_0}(x^2-1)\left(1-\frac{p_0^2+r_0^2-2p^2-2m_{_{F}}^2}{2pq_0}\right)
  \beta_{-}(P)\Delta_{-}(R)\right\}
\label{eq:countself}
\end{eqnarray}
where $x\equiv{p_0}/{p}$ and ${\partial}/{\partial p}$ is the partial
derivative with respect to $p$ at constant $x$, and where
$\Delta_{\pm}(P)$ is the time-like part of the spectral density in
Eq.~(\ref{eq:spectfer}), which can be approximated to logarithmic
accuracy by $\delta(p_0\mp\omega_{+}(p))$ ({\it i.e.\/} the residue of
the pole is approximated by $1$, which is its value at large momentum).

Finally, to logarithmic accuracy, we get:
\begin{equation}
 \left. \frac{dN}{d^4xdq_0 d^3{\imb q}}\right|_{\rm self}\approx\frac{N}{12\pi^4}\Big(\sum_f e^2_f\Big)\frac{\alpha^2 m_{_{F}}^2}{q_0^2}\ln\left(\frac{q_0T}
  {q_0^2+m_{_{F}}^2}\right)\; .
\label{eq:logcount}
\end{equation}

\bibliographystyle{unsrt}

\end{document}